\documentclass[twocolumn]{aastex631}

\newenvironment{equationc}{\begin{equation}}{,\end{equation}\ignorespacesafterend}

\graphicspath{{./}{figures/}}
\usepackage{parskip}
\usepackage{gensymb}
\usepackage{subfigure}
\usepackage{graphicx}
\usepackage{amssymb, mathtools, makecell}
\usepackage{tabularx} 
\usepackage{xcolor}

\begin{document}
\title{Prospects for extending the core-collapse supernova detection horizon using high-energy neutrinos}

\author{Nora Valtonen-Mattila}
\email{nora.valtonen@physics.uu.se}
\author{Erin O'Sullivan} \email{erin.osullivan@physics.uu.se}
\affil{Department of Physics \& Astronomy, Uppsala University}

\begin{abstract}

Large neutrino detectors like IceCube monitor for core-collapse supernovae using low energy (MeV) neutrinos, with a reach to a supernova neutrino burst to the Magellanic Cloud. However, some models predict the emission of high energy neutrinos (GeV-TeV) from core-collapse supernovae through the interaction of ejecta with circumstellar material and (TeV-PeV) through choked jets. In this paper, we explore the detection horizon of IceCube for core-collapse supernovae using high-energy neutrinos from these models. We examine the potential of two high-energy neutrino data samples from IceCube, one that performs best in the northern sky and one that has better sensitivity in the southern sky. We demonstrate that by using high-energy neutrinos from core-collapse supernovae, the detection reach can be extended to the Mpc range, far beyond what is accessible through low-energy neutrinos. Looking ahead to IceCube-Gen2, this reach will be extended considerably.

\end{abstract}


\section{Introduction} \label{sec:intro}

The core-collapse supernovae (CCSNe) explosion mechanism is driven by low energy (MeV) neutrinos, which are responsible for releasing most of the gravitational binding energy of the system. These neutrinos were observed for the first time in 1987 by Kamiokande-II \citep{Hirata:1987_12}, Irvine-Michigan-Brookhaven detector \citep{Bionta:1987_13}, and Baksan \citep{Alexeyev:2008_32}, where 24 candidate neutrino events were observed between the three detectors. In addition to MeV neutrinos, some CCSNe are good candidates for the production of high-energy (HE) neutrinos GeV and higher due to dense circumstellar material (CSM), which provides target material for the ejecta to form shocks and accelerate protons with matter via hadronuclear ($pp$ interaction) or photohadronic ($p\gamma$) mechanism. While these neutrinos have yet to be observed, there are recent hints of HE neutrinos in connection with the SN 1987A \citep{Oyama:2022_37}.

The expected rate of a Galactic CCSN is $\sim $ a few per century \citep{Rozwadowska:2021_19, Adams:2013_46}, therefore considerable work has been done in understanding the detector capability of observing MeV neutrinos in anticipation of the next event. Neutrino telescopes such as the IceCube Neutrino Observatory \citep{IceCube:2016_42} and KM3NeT \citep{KM3Net:2016_43} observe the burst of MeV neutrinos from CCSNe as single-photon hits from interactions that occur near the photomultiplier tubes (PMTs). This makes the measurement quite sensitive to the high level of single photon noise due to, for example, the dark noise in the PMT or radioactivity of the PMT glass. This limits the ability for neutrino telescopes to observe faint signals from CCSNe and so, despite the large size of the detector, IceCube has a detection horizon for CCSNe of $\sim 50$ kpc \citep{IceCube:2011_33}, and KM3NeT has an expected reach of $\sim 50-60$ kpc \citep{KM3NeT:2021_22}, in contrast to detectors with tighter PMT spacing like Super-K, which has nearly a double reach of $\sim 100$ kpc \citep{Ikeda:2007_20}. For next-generation detectors like Hyper-K, the detection horizon for low energy neutrinos from CCSNe is expected to reach $\sim 1$ Mpc \citep{Nakamura:2016_28}.

In this paper, we explore the potential to extend the detection horizon past the Magellanic Clouds through the detection of HE neutrinos. In addition to increasing the number of observable supernovae, these neutrinos could give us an insight into the cosmic ray acceleration processes, which are inaccessible with the low-energy neutrinos produced through nuclear processes in supernovae. We consider two production mechanisms for HE neutrinos in CCSNe: One through the interaction between supernova ejecta with CSM and the other through relativistic choked jets.

The CSM-ejecta model was first proposed in \citet{Murase:2011_47} and predicted neutrino emission times between 0.1 day to 1-year post-core bounce. This model was recently extended in \citet{Murase:2018_01} and \citet{Kheirandish:2022_44}, and this HE neutrino flux can contribute to the flux of diffuse neutrinos found by IceCube \citep{IceCube:2015_30, Necker:2021_31, Zirakashvili&Ptuskin:2016_25}. Key parameters affecting the HE neutrino flux include mass loss, wind velocity, shock velocity, and proton spectral index, as investigated in \citet{Sarmah&Tamborra:2022_45} and \citet{Murase&Franckowiak:2019_57}.

In the choked jet (CJ) scenario, the proposed progenitors are similar to gamma-ray bursts (GRBs), except the jets are slower and never break through the stellar envelope, leading to neutrinos without the counterpart gamma-ray emission \citep{Senno&Murase:2016_26}. This scenario was initially proposed by Razzaque, M\'esz\'aros, and Waxman (RMW) in \citet{Razzaque&Meszaros&Waxman:2004_18,Razzaque&Meszaros&Waxman:2005_10}, with an emission time from 10 s \citep{Razzaque&Meszaros&Waxman:2004_18, Shinichiro&Beacom:2005_50, Enberg:2009_02,Bromberg&Nakar&Piran:2011_62} to $10^4$ s \citep{Murase:2013_61}, with several parameters affecting the neutrino flux, such as the Lorentz bulk factor $\Gamma_b$ \citep{IceCube_SN:2011_52}, injected energy, and engine time \citep{He:2018_60}. In the slow jet CJ scenario used in this work, the $\Gamma_b$ is estimated to be $\lesssim 10$ \citep{Razzaque&Meszaros&Waxman:2004_18,Razzaque&Meszaros&Waxman:2005_10,Enberg:2009_02} compared to GRBs with $\gtrsim 100$ \citep{Razzaque&Meszaros&Waxman:2004_18}. The RMW model \citep{Razzaque&Meszaros&Waxman:2004_18} was expanded in the work of \citet{Shinichiro&Beacom:2005_50}, where they explored the kaon component. A charm meson contribution to the flux was added in \citet{Enberg:2009_02}, and the contribution to the diffuse neutrino flux was investigated in \citet{Bhattacharya&Enberg:2015_07}. Other models have been proposed for a similar choked scenario, such as in \citet{Murase:2006_51}, where they investigated the neutrino flux for higher $\Gamma_b$ or in \citet{Murase:2013_54} where they demonstrated that TeV-PeV neutrinos are possible from these sources. These sources can also contribute to IceCube's observed flux \citep{Senno&Murase:2018_55, Esmaili&Murase:2018_56, IceCube_SN:2011_52, IceCube_SN:2012_53}. Finally, a framework for the optical follow-up of HE neutrino transients is currently running with telescopes such as the Zwicky Transient Facility (ZTF) \citep{Nordin:2019_29}.

Building on this work, we investigate the detection horizon for different types of CCSNe for IceCube, using models that provide moderate neutrino flux predictions and expand on previous work. This work can be applied to near-future cubic-kilometer-scale detectors like KM3NeT, Baikal-GVD \citep{Avronin:2019_39}, and P-ONE \citep{P-ONE:2020_40}, which will have a similar sensitivity in the southern sky as IceCube has to the northern sky. In section 2, we will introduce the different HE neutrino production models from CCSNe and discuss the main background sources in the IceCube detector. We will then demonstrate the procedure for determining the detection horizon in section 3, show our results in section 4, and discuss the relevance of our results in section 5.

\section{HE supernova neutrino signal} \label{sec: Background}

\subsection{Production mechanisms for HE neutrinos in CCSNe} \label{subsec:high-E}
In this paper, we consider two models that predict the production of HE neutrinos via two mechanisms: ejecta interaction with CSM from \citet{Murase:2018_01} and relativistic CJ from \citet{Enberg:2009_02}.

Supernovae that have experienced mass loss before the explosion due to, for example, stellar winds are surrounded by CSM. The type of supernovae is characterized by optical observations and are associated with different amounts of CSM. Type IIn is associated with having a significant amount of CSM in the $O$($10^{-3}-10^{-1}$) M$_\odot$yr$^{-1}$ \citep{Moriya:2014_48}, and type II-P have the potential for significant CSM in the $O$($10^{-3}-10^{-2}$) M$_\odot$yr$^{-1}$ \citep{Murase:2018_01, Moriya:2018_49}, making them good candidates for HE neutrino emission via Fermi shocks. Type Ib/c and IIb are candidates for HE neutrino emission from CJ \citep{Piran:2019_11}.

\textbf{CSM-Ejecta Interaction}

When the supernova explodes, the ejecta compresses the CSM, forming shocks. These shocks propagate in the CSM, creating an environment where charged particles are trapped and scattered. Via inelastic pp collisions, HE neutrinos are produced through processes like
\[pp\rightarrow\pi^+ \rightarrow \mu^+ \nu_\mu \rightarrow e^+ \nu_{e}\nu_\mu \bar{\nu_\mu}\] .

The predicted neutrino flux depends on the CCSNe explosion parameters, including explosion energy, wind velocity, ejecta mass, and mass-loss rate. For this work, we assume the model from \citet{Murase:2018_01}, where a time-dependent neutrino flux is obtained based on semi-analytic modeling of particle acceleration in a dense CSM.

\textbf{CJ}

We also consider a model where the production of HE core-collapse supernova neutrinos arises via CJ \citep{Enberg:2009_02}. In this scenario, a mildly-relativistic jet in the collapsing star becomes trapped behind the optically-thick outer shell. In contrast to gamma-ray bursts, where gamma rays produced in the jet can escape the star, only neutrinos escape the star in the CJ scenario. To predict the neutrino flux produced in CJ, we use the model described in \citet{Enberg:2009_02}, which includes the decay of charm mesons in addition to pion and kaon decay.

\subsection{Detection in IceCube} \label{sec:IceCube}

IceCube observes HE neutrinos via Cherenkov radiation due to charged particles produced when neutrinos interact with nucleons in the ice. The resulting topology can be track-like or cascade-like, depending on the final state. Track-like events are created by a final state $\mu$ that can propagate long distances. Due to the long track of photons produced in the detector, this topology is ideal for the directional reconstruction of the neutrino, with an angular resolution of $1\degree$ or better, but is more challenging for energy reconstruction as the muon can deposit energy outside of the instrumented volume. Cascade-like events, have a shorter length than track-like events and more of the energy is contained within the detector. This event topology is ideal for energy reconstruction but has a large directional uncertainty of $\gtrapprox10\degree$.

In the IceCube detector, the main background to the astrophysical signal are the muons and neutrinos produced when HE cosmic rays interact in the Earth's atmosphere. For the northern sky (declination $\delta > -5\degree$), the Earth strongly attenuates the atmospheric muons and so the dominant remaining background is atmospheric neutrinos. In the energy range of interest, for the southern sky ($\delta < -5\degree$), the background is significantly higher as removing the atmospheric muons is a challenge since there is no filtering through the Earth.

\section{ANALYSIS} \label{sec:Analysis}
\subsection{Determining the number of observable neutrinos} \label{subsec:flux}

To calculate the number of HE core-collapse neutrinos observable in IceCube, we consider two different data samples: the IceCube 10-year data release \citep{IceCube:2021_05}, which contains neutrino-induced track-like muon events with good sensitivity in the northern sky, and the Medium Energy Starting Events (MESE) \citep{IceCube:2015_06} data sample, which consists mostly of cascades and achieves a better sensitivity in the Southern sky where the atmospheric muon background is high by selecting events with the vertex inside of the instrumented volume.

To obtain the mean number of neutrinos $N_\nu$ observable by IceCube, we convolve the neutrino flux at Earth $\phi(E_\nu,t)$ with the declination-averaged effective area $A_{eff}(E_\nu)$ and then integrate over time and energy such that

\begin{equationc} \label{eqn:n_nu}
    N_\nu=\frac{1}{(d/d_{ref})^2}\int_{t_{min}}^{t_{max}} \int_{E_{min}}^{E_{max}} \phi(E_\nu,t)\, A_{eff}(E_\nu)\,dt\,dE_\nu
\end{equationc}

where $E_{min}$, $E_{max}$, $t_{min}$, $t_{max}$, are the minimum and maximum neutrino energy and the minimum and maximum time of observation for each model. The number of neutrinos is scaled as a function of distance $d$ against the $d_{ref}$ which is the reference distance for each model's flux. The effective area is a parametrization of detector sensitivity provided in IceCube data samples \citep{IceCube:2021_05,IceCube:2015_06} that depends on neutrino energy, flavor, neutrino direction, and absorption effects due to Earth. We averaged the effective area over each hemisphere, with the northern hemisphere defined as $\delta > -5^{\circ} $ and southern hemisphere as $\delta < -5^{\circ} $.

\textbf{CSM-ejecta interaction model}

For the model of CCSNe ejecta interacting with CSM, we use \citet{Murase:2018_01}, choosing the supernova types with the two largest fluences: type IIn and II-P, and from those, the 'optimistic' neutrino fluxes with a proton momentum index of 2.0 are chosen. We assume a ratio of $\nu_\mu :\nu_e : \nu_\tau \approx 1:1:1$ at Earth, and we integrate from $t_{min}=10^3$ s for II-P and $t_{min}=10^5$ s for IIn. These minimum observation times represent the time of onset of the CR acceleration, relative post-core bounce, as specified in \citet{Murase:2018_01}. The onset time for CR acceleration is proportional to the ejecta velocity, mass loss rate, ejecta mass, and kinetic energy, being different for type IIn and II-P. For the maximum time of integration, we use $t_{max} = 10^{5.8}$ s for type II-P and $t_{max} = 10^{7}$ s for type IIn based on the model emission time from \citet{Murase:2018_01}. The reference distance for this model given in \cite{Murase:2018_01} is $d_{ref}=10$ kpc.

\textbf{CJ model}

For the CJ model, we use the fluxes from \citet{Enberg:2009_02} with $pp\rightarrow D^\pm$, $pp/p\gamma \rightarrow K^\pm$, $pp / p\gamma \rightarrow \pi^\pm$,  channels. For $pp\rightarrow D^\pm$ channel, we assume a $\nu_e :\nu_\mu : \nu_\tau \approx 1:1:0$ at the source, for $pp/p\gamma \rightarrow K^\pm$, we assume $\nu_e :\nu_\mu : \nu_\tau \approx 1:2:0$, and for $pp / p\gamma \rightarrow \pi^\pm$, we assume $\nu_e :\nu_\mu : \nu_\tau \approx 1:2:0$ \citep{Kachelriess:2006_27} at source, and after mixing, we assume all fluxes to be $\nu_e :\nu_\mu : \nu_\tau \approx 1:1:1$ at Earth. The onset time is assumed to be $t_{min}=0$ and the maximum is taken to be the duration of the burst at $t_{max}=10$ s \citep{Bhattacharya&Enberg:2015_07,Koers:2007_08}. We then sum each of the individual decay contributions to obtain a total flux. The reference distance for this model, as given in \citet{Enberg:2009_02}, is $d_{ref}=20$ Mpc.

\section{Results} \label{sec:Results}

\subsection{CSM-ejecta interaction model} \label{subsec:ejecta_CSM}

\begin{figure}
\centering     
\subfigure[Northern sky]{\label{fig:a}\includegraphics[width=0.48\textwidth]{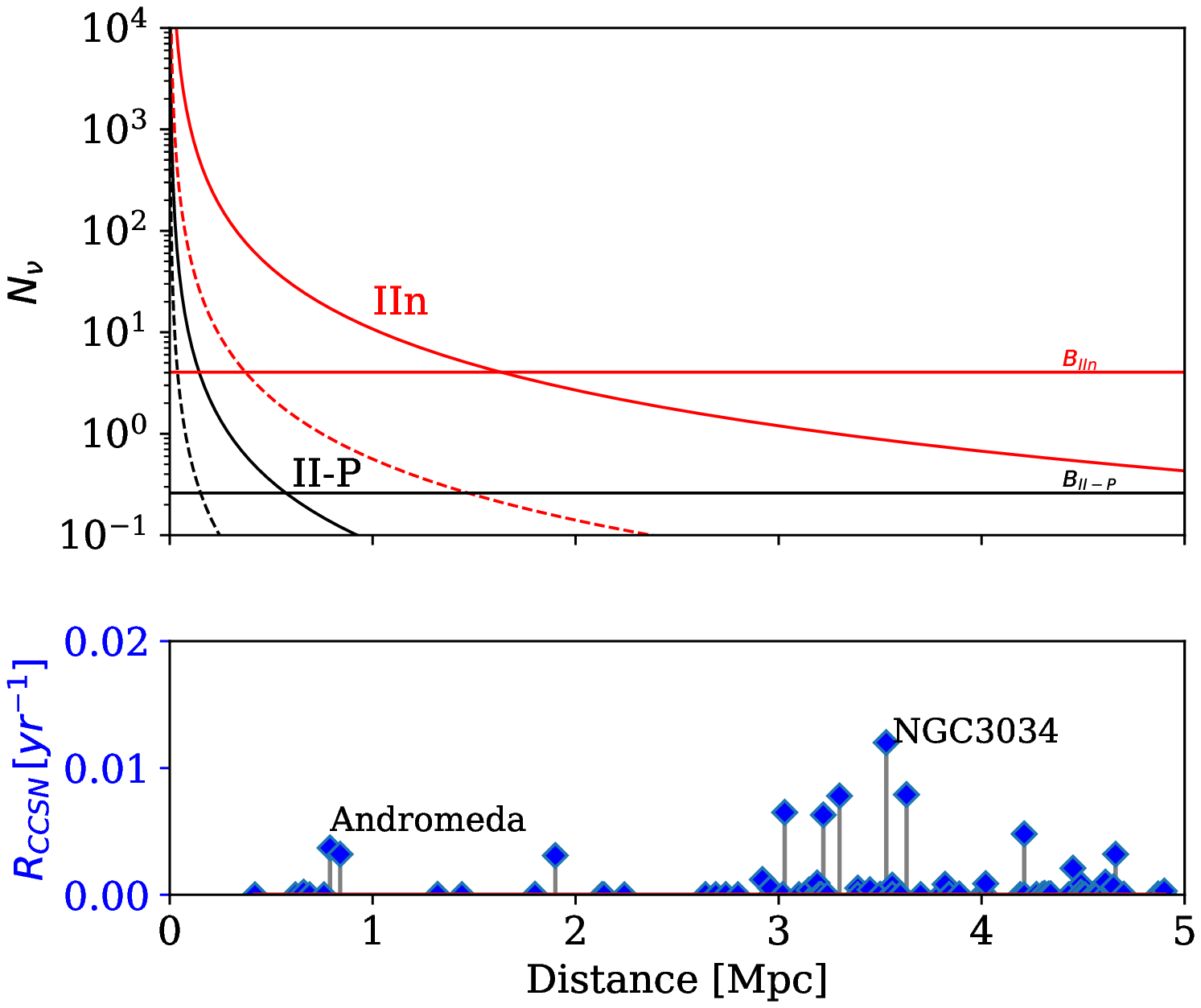}}
\subfigure[Southern sky]{\label{fig:b}\includegraphics[width=0.48\textwidth]{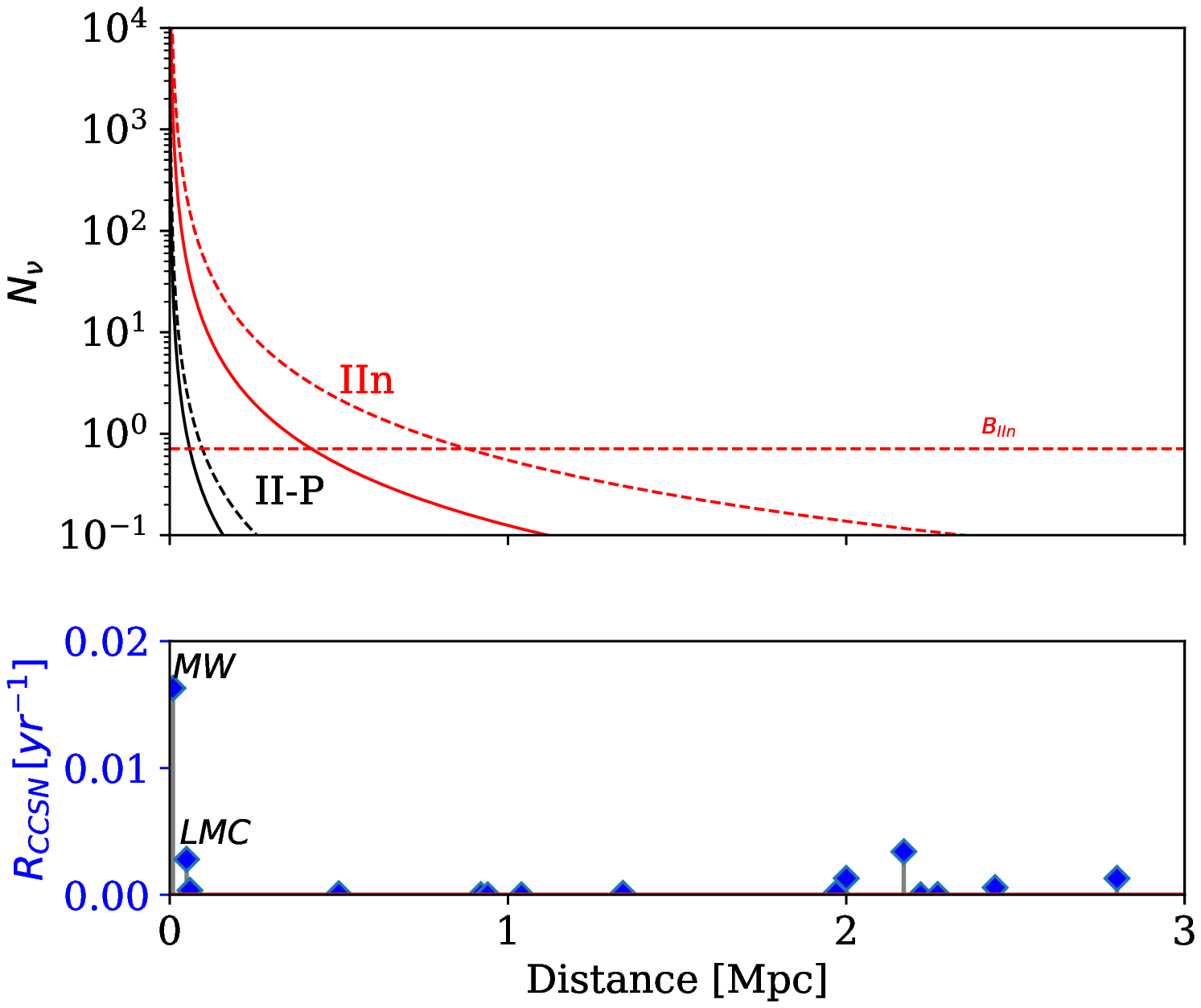}}
\caption{Number of observable neutrinos by IceCube for CSM-ejecta model using tracks (solid) and MESE (dashed), for type II-P (black) and type IIn (red) together with the yearly CCSNe rate from \citet{Nakamura:2016_28}. Fig. a) Upper panel shows the expected number of neutrinos observable by IceCube for the northern sky using tracks and MESE as a function of distance with the background rate for tracks indicated as solid horizontal line, together with the yearly CCSNe rate (lower panel). Here Andromeda (M 31) labeled for reference and NGC 3034 is the galaxy with the highest CCSN rate in the northern sky. Fig. b) Expected number of neutrinos observable by IceCube for the southern sky using tracks and MESE for CCSNe type II-P and IIn, with the background rate for MESE indicated as solid horizontal line together with the yearly CCSNe rate (lower panel), with the Milky Way (MW) and LMC labeled for reference.}
\label{fig:tracks}
\end{figure}

We evaluate the number of neutrinos observed in IceCube under the assumptions of the CSM-ejecta model individually for the northern and southern hemispheres using equation \ref{eqn:n_nu}. We find that for a type II-P (IIn) SN, which accounts for 52.5\% (6.75\%) of all CCSNe \citep{Li:2011_09}, IceCube has a detection horizon for muon tracks in the northern sky of 0.2 (2.3) Mpc for a doublet and 0.3 (3.3) Mpc for a single neutrino. In the southern sky, the sensitivity using the MESE data sample can be extended to 0.06 (0.5) Mpc for a doublet and  0.08 (0.74) Mpc for a single neutrino.

Figure \ref{fig:tracks} shows the number of neutrinos expected for the CSM-ejecta model in IceCube. The number of background events we expect for track-like events in the northern sky is estimated to be 0.26 for the integration time of type II-P and 4.1 for the integration time type IIn, assuming an average circularized angular uncertainty 1$\degree$. For the MESE sample in the southern sky, the expected background is 0.04 for type II-P and 0.71 for IIn assuming a circularized average angular uncertainty of 10$\degree$. Figure \ref{fig:tracks} also shows the CCSN rate from \citet{Nakamura:2016_28}, which used the H$_\alpha$ line from each galaxy corrected for [NII] line contamination and Galactic extinction to determine the rate. We used the Galactic CCSNe rate estimate from \citet{Rozwadowska:2021_19}.

\subsection{CJ HE neutrino production} \label{subsec:ejecta_charm}

For the CJ model, the reach is extended considerably more than that of the CSM-ejecta model. The singlet (doublet) detection horizon is 85(60) Mpc for the northern sky using tracks and 20 (14) Mpc for the southern sky using the MESE selection. 

\begin{figure}
\centering     
\subfigure[Northern sky]{\label{fig 2:a}\includegraphics[width=0.49\textwidth]{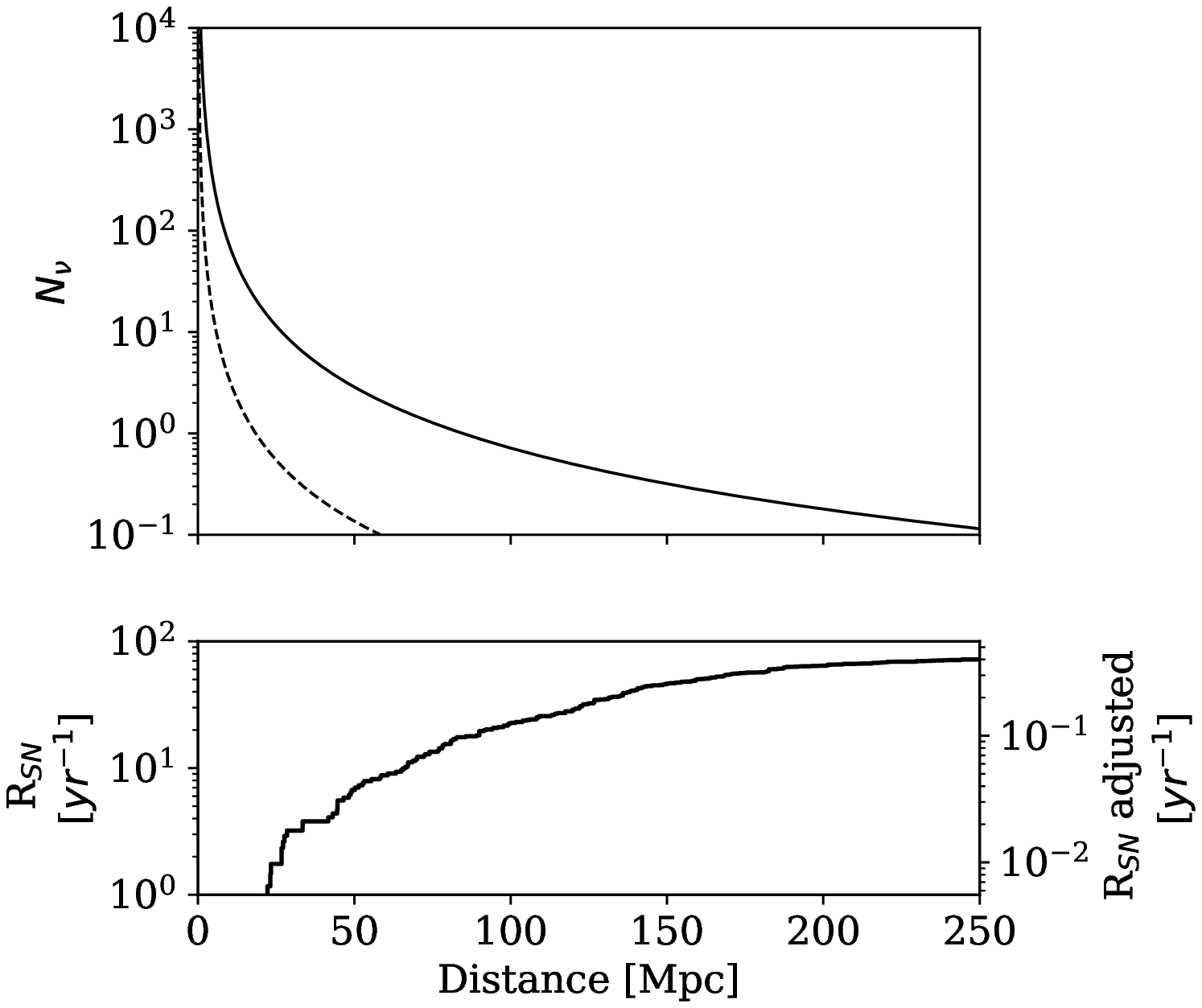}}
\subfigure[Southern sky]{\label{fig 2:b}\includegraphics[width=0.49\textwidth]{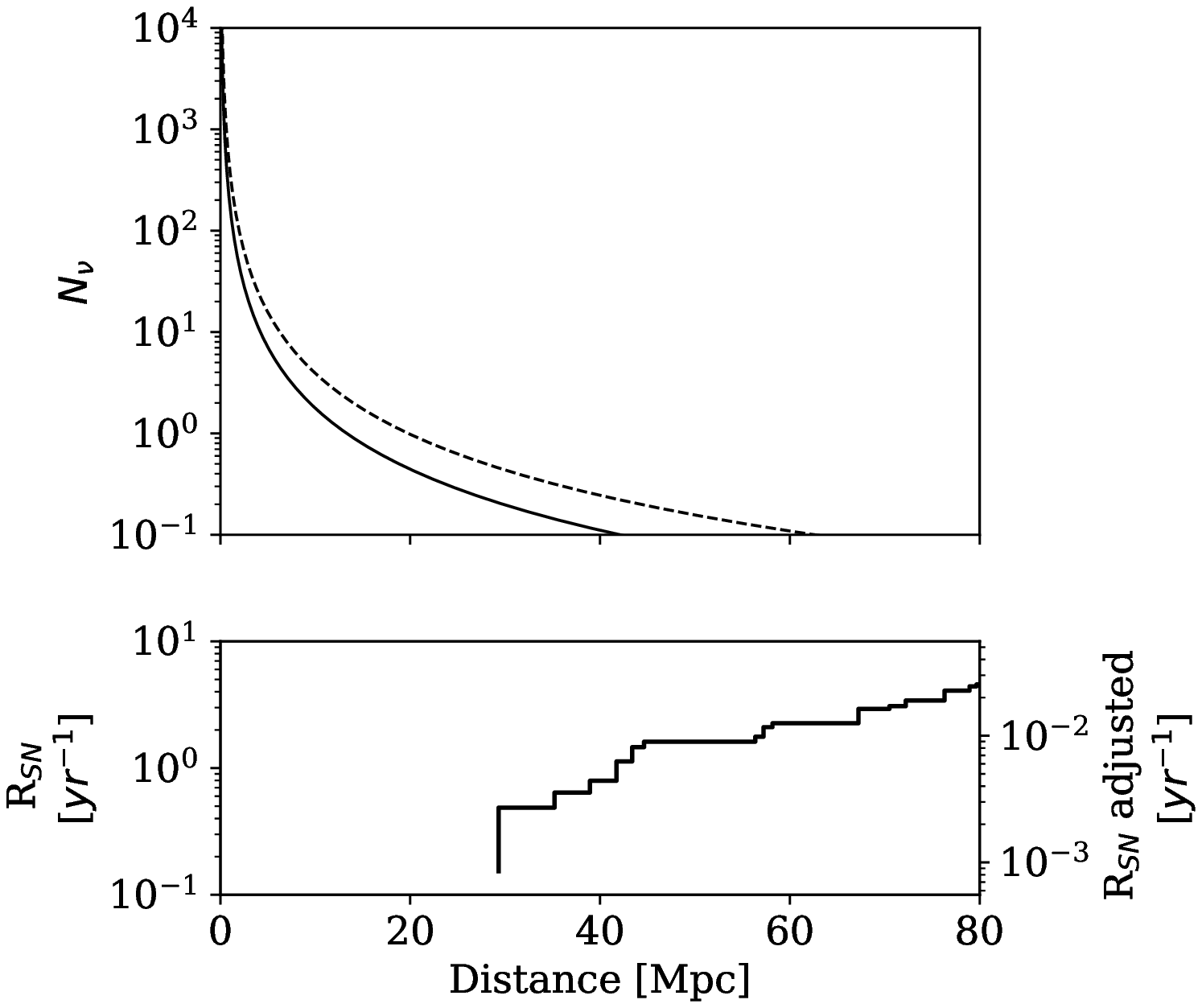}}
\caption{Number of observable neutrinos by IceCube using CJ model for tracks (solid line) and MESE (dashed line) Fig. a) The upper panel shows the expected number of neutrinos observable by IceCube for the northern sky and the lower panel shows the ZTF yearly cumulative observation of type Ib/c + IIb supernovae \citep{ZTF:2020_41}, with the left y-axis showing the yearly observed denoted as R$_{SN}$, and the right y-axis the adjusted yearly rate observable through neutrinos after a suppression factor of $1/180$. Fig. b) The upper panel shows the expected number of neutrinos observable by IceCube and the lower panel shows the cumulative observed through ASAS-SN \citep{ASASSN:2014_62} and ZTF, where the left y-axis showing the yearly observed and the right y-axis the adjusted yearly rate.} 
\end{figure}

Figure \ref{fig 2:a} shows the number of neutrinos expected to be observed in IceCube for the CJ model, together with nearby, electromagnetically-observed CCSNe type Ib/c and IIb. We use the ZTF catalog\footnote{\url{https://sites.astro.caltech.edu/ztf/bts/bts.php}} \citep{ZTF:2020_41} for the northern sky and a combined catalog of sources from ZTF and ASAS-SN \citep{ASASSN:2014_62} \footnote{\url{https://www.astronomy.ohio-state.edu/asassn/index.shtml}} for the southern sky.

Given the short emission time of 10 s for the CJ model, the expected background rate for a singlet is in $O$($10^{-6}$)($O$($10^{-7}$)) neutrinos for the northern (southern) sky. In order to determine that a nearby supernova of interest has occurred, a coincident electromagnetic observation is required. The mean uncertainty for the explosion time for type Ib/c supernovae is 13 days \citep{Cano:2017_59, Esmaili&Murase:2018_56,Senno&Murase:2018_55}. The expected background rate for one neutrino candidate within 13 days and from the direction of the supernova is $O$($10^{-1}$) ($O$($10^{-2}$)). The expected background rate for a doublet arriving within a 10 s window and from the direction of the supernova in those 13 days is $O$($10^{-2}$) ($O$($10^{-6}$)) \footnote{The number of doublets is considerably smaller in the southern sky due to a low number of accepted events in the MESE data sample compared to the tracks data sample.}.

Table \ref{tab:galaxies} summarizes the number of observable neutrinos $N_\nu$ for the top 20 galaxies within 5 Mpc for both models.

This study uses a declination-averaged effective area $A_{eff}(E_\nu)$. The neutrino energy range predicted from the models is sufficiently low that the sensitivity is not significantly impacted by direction. For track-like events in the northern sky, the detection horizon differs by $\sim$ + 7$\%$ /-- 19$\%$ depending on the declination of the incoming neutrino. For the MESE selection cut in the southern sky, the detection horizon varies by $\sim$ + 5$\%$ /-- 3$\%$. 

\begin{deluxetable*}{cccccccc}
\tablenum{1}
\tablecaption{Top 20 galaxies\label{tab:galaxies}}
\tablewidth{2pt}
\tablehead{
\colhead{Galaxy} & \colhead{RA} & \colhead{Dec} & \colhead{Distance} & \colhead{CCSN Rate}& \colhead{$N_\nu$[II-P]} & \colhead{$N_\nu$[IIn]} & \colhead{$N_\nu$[Choked jets]}\\
\colhead{Name} & \colhead{(Deg)} & \colhead{(Deg)} & \colhead{(Mpc)} & \colhead{(yr$^{-1}$)} &  \colhead{Number} & \colhead{Number} & \colhead{Number}
} 
\decimalcolnumbers
\startdata
NGC 5236 & 204.25 & -29.87 & 4.47 & 0.0240 & 0.0003 & 0.028 & 19.6 \\
NGC 3034 & 148.97 & 69.68 & 3.53 & 0.0120 & 0.0069 & 0.86 & 575 \\
NGC 253 & 11.89 & -25.29 & 3.94 & 0.0120 & 0.0004 & 0.0353 &  25\\
NGC 5128 & 201.37 & -43.02 & 3.66 & 0.0092 & 0.0005 & 0.041 & 29 \\
NGC 3031 & 148.89 & 69.07 & 3.63 & 0.0079 & 0.0065 & 0.82 & 544 \\
Maffei 2 & 40.48 & 59.60 & 3.30 & 0.0078 & 0.008 & 1 & 658 \\
UGC 2847 & 56.70 & 68.09 & 3.03 & 0.0065 & 0.009 & 1.17 & 780 \\
NGC 4945 & 196.37 & -49.47 & 3.60 & 0.0064 &  0.0005 & 0.042 & 30\\
NGC 2403 & 114.21 & 65.60 & 3.22 & 0.0063 & 0.008 & 1.04 & 691 \\
NGC 4449 & 187.05 & 44.09 & 4.21 & 0.0048 & 0.005 & 0.60 & 404 \\
NGC 1313 & 49.57 & -66.49 & 4.47 & 0.0044 & 0.0004 & 0.032 & 23 \\
M 31 & 10.69 & 41.27 & 0.79 & 0.0037 & 0.137 & 17.2 & 1.15$\cdot 10^4$ \\
NGC 7793 & 359.46 & -32.59 & 3.90 & 0.0037 & 0.0004 & 0.036 & 26 \\
NGC 55 & 3.73 & -39.19 & 2.17 & 0.0034 & 0.0013 & 0.117 & 83 \\
NGC 598 & 23.46 & 30.66 & 0.84 & 0.0032 & 0.12 & 15.3 & 1.02$\cdot 10^4$ \\
NGC 4736 & 192.72 & 41.12 & 4.66 & 0.0032 & 0.004 & 0.4955 & 330 \\
NGC 1569 & 67.70 & 64.85 & 1.90 & 0.0031 & 0.024 & 2.98 & 2$\cdot 10^3$ \\
LMC & 80.89 & -69.76 & 0.05 & 0.0028 & 2.65 & 219.5 & 2.87$\cdot 10^6$ \\
NGC 4236 & 184.18 & 69.46 & 4.45 & 0.0021 & 0.004 & 0.543 & 362 \\
NGC 247 & 11.79 & -20.76 & 3.65 & 0.0020 & 0.0005 & 0.041 & 29 \\
\enddata
\tablecomments{This table shows the top 20 galaxies that comprise 87\% of all the CCSN rate within 5 Mpc from \citet{Nakamura:2016_28}}
\end{deluxetable*}

\section{DISCUSSION} \label{sec:Discussion}

We obtained the sensitivity of IceCube for HE neutrinos from CCSNe for the northern and southern hemispheres, using two different models for HE neutrino production in CCSNe. We consider type II-P and IIn for the CSM-ejecta model and type Ib/c and IIb for the CJ model. The types of CCSNe considered in this work consist of $\sim 87\%$ of all CCSNe. Our results show that the reach of the IceCube detector for CCSNe can be extended past the LMC using HE neutrinos from the CSM-ejecta mechanism and CJ. 

Type II-P is the most common type of CCSNe, accounting for 52.5\% of all CCSNe. The reach in the northern sky extends to 300 kpc for 1 neutrino. For the southern sky, the detection horizon for 1 neutrino extends past the LMC to 80 kpc. However, the galaxies within this expanded detection volume consist mostly of small dwarf galaxies, where the rate of CCSNe is low. None of the galaxies in table 1 are on the detection horizon for type II-P.

For Type IIn and CJ candidates (Ib/c and IIb) the detection horizon extends to the Mpc scale for the northern sky. For type IIn, which consists of $\sim 7\%$ of all CCSNe, the expected number of observed tracks for a CCSN in Andromeda is 17 neutrinos, and 15 neutrinos for nearby NGC598. The single neutrino detection horizon in the northern sky is 3.3 Mpc (see Figure 1a and Table I), reaching the region where Centaurus A and M81 cluster reside which provide a high density of galaxies nearby \citep{Nakamura:2016_28}. For CJ, which consists of $\sim 27\%$ of all CCSNe, the reach extends to tens of Mpc; in the northern sky, using tracks, we can reach all of the top 20 galaxies (Table 1). The detection horizon for the CJ model extends to the range where telescopes such as ZTF have observed CCSNe (Figure 2a). Since the background is negligible for the CJ model due to the short duration of the burst, even a single neutrino connected to an optically observed supernova would be significant. The detection horizon is 85 Mpc for a single neutrino and 60 Mpc for a doublet.

For the southern sky, for type IIn and using the MESE selection cut, the detection horizon can be extended past the LMC and that of type II-P. We expect 220 neutrinos to be observed in IceCube from the LMC, with a detection horizon of 520 kpc for a doublet and 740 kpc for a singlet (Figure 1b). With CJ, we can extend the detection horizon to all of the top 20 galaxies (Table 1) and reach 20 Mpc with 1 neutrino and 63 Mpc with 10\% probability of observing a neutrino (fig. 2b). The CJ model can extend the detection horizon to confirmed observations by ZTF and ASAS-SN. In this work, we made use of a modest prediction model, and it is important to note that there are uncertainties in many parameters that can influence the resulting neutrino flux. For the CSM-ejecta model, for example, the shock velocity and proton energy can influence the detection horizon by one order of magnitude, as demonstrated in \citet{Sarmah&Tamborra:2022_45}, where they give a reach for a type II-P (IIn) of 0.2-2 (0.6-6) Mpc for IceCube. As shown in \citet{Murase&Franckowiak:2019_57}, the neutrino fluence is also sensitive to the proton index s. We used of the more optimistic estimate from \citet{Murase:2018_01}, with s = 2.0, which is expected for quasi-parallel shocks \citep{Murase&Franckowiak:2019_57}. However, for larger s, the neutrino flux is expected to decrease \citep{Murase&Franckowiak:2019_57,Murase:2018_01}. There are other models, such as in \citet{Kheirandish:2022_44}, where they demonstrate that the neutrino fluence is also sensitive to the density of the CSM, proposing even more optimistic prospects for type II-P than presented in this work. The CJ model also has uncertainties, for example, in the $\Gamma_b$, injected energy, and engine time. For example, variation in the $\Gamma_b$ can affect the event rates observable by IceCube \citep{IceCube_SN:2012_53}, where for typical energy of $E \sim 10^{51}$ erg, a larger $\Gamma_b$ could increase the event rate observable by $O$($10^2$) and a smaller $\Gamma_b$ could decrease the event rate by $O$($10^1$). The CJ model used in this work assumes a $\Gamma_b = 3$; however slow CJ can present all the way to $\Gamma_b \lesssim 10$, with the potential to extend the detection horizon past the one presented in this work. Similarly, in \citet{He:2018_60}, they predict the triplet IceCube detection horizon to be at 81-600 Mpc, highlighting how the parameter space can improve the prospects for detection.

Although the prospects with CJ candidates are promising, this type of HE neutrino mechanism is rare compared to the CSM-ejecta scenario, with estimates ranging from 1-4\% of all CCSNe \citep{Shinichiro&Beacom:2005_50,IceCube_SN:2012_53,Razzaque&Meszaros&Waxman:2005_10,Piran:2019_11}. In this work, we assumed a moderate prediction with 10\% of all Ib/Ic and potentially IIb \citep{Razzaque&Meszaros&Waxman:2005_10,Piran:2019_11} or equivalently 2.7\% of all CCSNe. In addition, there is also a suppression factor that arises from only $1/(2\Gamma_b^2)$ having their jets pointed at Earth \citep{Razzaque&Meszaros&Waxman:2005_10}. For the model used in this work \citep{Enberg:2009_02}, with a $\Gamma_b = 3$, if we consider a moderate prediction of 10\% of all Ib/Ic and IIb, we would obtain $1/180$ supernovae with slow jets pointing at Earth. The lower panels of fig. 2a and 2b show how this suppression factor would scale the optically detected type Ib/c + IIb, expressed as R$_{SN}$, and the adjusted rate with the suppression factor as R$_{SN}$ adjusted. For the northern sky, at a doublet (singlet) distance of 60 (85) Mpc, there were 10 (17) observed type Ib/c + IIb per year. With the suppression factor, we would expect $\sim$0.06 (0.1) observable supernova candidates with at least two (one) neutrino per year or one observation through neutrinos connected to CJ emission in $\sim$15 (10) years in IceCube. However, this number will vary depending on the assumption about the population that could harbor jets. If only 1\% of the population has slow jets directed at Earth, we expect $\sim$0.14 (0.35) supernovae in 10 years of IceCube data for the doublet (singlet) horizon or up to $\sim$0.5 (1.5) if 4\% of the CCSNe population has CJ. Since we cannot differentiate between the type Ib/c+IIb that would have jets from those that don't have jets, in an neutrino-optical coincident search we would use a full catalog of nearby type Ib/c+IIb. However, a doublet coincident with an optical detection would still be significant, even with the penalty factor, and would give insight into potential jet acceleration mechanisms.

The sensitivities presented here are for the current IceCube detector. IceCube has better sensitivity with track-like events for the northern sky, with a directional uncertainty of 1$\degree$ or better, which allows us to identify the source with good accuracy. However, the track-like sample does not perform well for the southern sky since the Earth is not filtering the background. The MESE selection cut, which contains both cascades and tracks, can improve the sensitivity in the southern sky. Although it has a worse angular resolution than the track sample ($\sim 10\degree$), it does allow for the detection of all flavors of neutrinos and not only $\nu_{\mu}$ tracks and provides an almost equal sensitivity for both hemispheres. With future improvements in the background reduction and the reconstruction of events, it is expected that the southern sky sensitivity will improve. In addition, when KM3NeT becomes fully operational, it is expected that their southern sky HE neutrino sensitivity will be better than IceCube, enabling an extended detection horizon in the southern sky. For IceCube Gen-2, where the detector volume is projected to increase by a factor of 10, one could expect an order of magnitude increase in the number of observed neutrinos from CSM-ejecta and CJ models, which would scale the detection horizon by a factor of 3. This increase in detection horizon for the CJ model would reach the range where ZTF has observed in the northern (southern) sky 250(9) type Ib/c+IIb, translating in 4 CCSNe observable in the northern sky through neutrinos in 10 years.

When Hyper-K becomes operational, it is expected to have a detection horizon of 1 Mpc for low energy neutrinos from CCSNe; however, using HE neutrinos from CJ and type IIn would still allow us to observe further away. This is important because the rate of CCSNe increases with observable volume. The rate based on optical observation is of 0.8 CCSNe per year for 5 Mpc and over 2 per year for over 10 Mpc \citep{Nakamura:2016_28}. This means that there are many CCSNe outside of the low-energy neutrino detection horizon but accessible through HE neutrinos. It is important to note that these HE neutrinos have yet to be observed, requiring specific conditions for acceleration to occur, whereas low-energy neutrinos are more certain to be emitted from CCSNe and have been observed. Currently, with these two models and this presented work, HE neutrinos are the only way to observe CCSNe past the LMC and into the range where robust observations are possible.

\section{CONCLUSIONS} \label{sec:Conclusions}

This work shows that the detection horizon of CCSNe using HE neutrinos can be extended past the LMC to the $O$(Mpc). Since the probability of observing CCSNe increases with observable volume, the capability of reaching further out increases our chances of observing the most powerful astrophysical explosion. We have demonstrated that HE neutrinos can significantly extend the detection horizon of CCSNe past the even the Mpc range in expected using low energy neutrinos in near-future neutrino detectors. 

Viewing HE neutrinos from a core-collapse supernova would be extremely significant, allowing us to probe the outer regions of the explosion and giving insight into the acceleration of particles in the surrounding material. IceCube already has potential to see these events and in the next generation IceCube-Gen2, there are firmer prospects. Preparing for the large number of HE neutrinos expected for a nearby CCSNe, and anticipating what will be possible with Gen2, will help ensure we are ready for these important phenomena.

\section{ACKNOWLEDGEMENTS} \label{sec: Acknowledgements}

The authors thank the members of the astroparticle physics group at Uppsala University and the IceCube group at Stockholm University, as well as Kohta Murase and Rikard Enberg, for their helpful discussions while preparing this work. We also acknowledge Segev BenZvi for comments on the draft of this manuscript. This work was supported by Vetenskapsr{\aa}det (the Swedish Research Council) under award numbers 2019-05447.

\newpage
\bibliography{references}

\begin{thebibliography}{}
\expandafter\ifx\csname natexlab\endcsname\relax\def\natexlab#1{#1}\fi
\providecommand{\url}[1]{\href{#1}{#1}}
\providecommand{\dodoi}[1]{doi:~\href{http://doi.org/#1}{\nolinkurl{#1}}}
\providecommand{\doeprint}[1]{\href{http://ascl.net/#1}{\nolinkurl{http://ascl.net/#1}}}
\providecommand{\doarXiv}[1]{\href{https://arxiv.org/abs/#1}{\nolinkurl{https://arxiv.org/abs/#1}}}

\bibitem[{{Aartsen} {et~al.}(2011)}]{IceCube:2011_33}
{Aartsen}, M.~G., {et~al.} 2011, \aap, 535, 18, \dodoi{https://doi.org/
  10.1051/0004-6361/201117810}

\bibitem[{{Aartsen} {et~al.}(2015{\natexlab{a}})}]{IceCube:2015_30}
---. 2015{\natexlab{a}}, \apj, 811, 52,
  \dodoi{https://doi.org/10.1088/0004-637X/811/1/52}

\bibitem[{{Aartsen} {et~al.}(2015{\natexlab{b}})}]{IceCube:2015_06}
---. 2015{\natexlab{b}}, \prd, 91,
  \dodoi{https://doi.org/10.1103/PhysRevD.91.022001}

\bibitem[{Aartsen {et~al.}(2017)}]{IceCube:2016_42}
Aartsen, M.~G., {et~al.} 2017, JINST, 12, P03012,
  \dodoi{10.1088/1748-0221/12/03/P03012}

\bibitem[{{Aartsen} {et~al.}(2021)}]{IceCube:2021_05}
{Aartsen}, M.~G., {et~al.} 2021, \dodoi{https://doi.org/10.21234/CPKQ-K003}

\bibitem[{Abbasi {et~al.}(2011)}]{IceCube_SN:2011_52}
Abbasi, R., {et~al.} 2011, Astronomy \& Astrophysics, 527,
  \dodoi{10.1051/0004-6361/201015770}

\bibitem[{Abbasi {et~al.}(2012)}]{IceCube_SN:2012_53}
---. 2012, Astronomy \& Astrophysics, 539, \dodoi{10.1051/0004-6361/201118071}

\bibitem[{Adams {et~al.}(2013)Adams, Kochanek, Beacom, Vagins, \&
  Stanek}]{Adams:2013_46}
Adams, S.~M., Kochanek, C., Beacom, J.~F., Vagins, M.~R., \& Stanek, K. 2013,
  ApJ, 778, \dodoi{10.1088/0004-637X/778/2/164}

\bibitem[{Adrian-Martinez {et~al.}(2016)}]{KM3Net:2016_43}
Adrian-Martinez, S., {et~al.} 2016, J. Phys. G, 43, 084001,
  \dodoi{10.1088/0954-3899/43/8/084001}

\bibitem[{{Agostini} {et~al.}(2020){Agostini}, {Böhmer}, {Bosma},
  {et~al.}}]{P-ONE:2020_40}
{Agostini}, M., {Böhmer}, M., {Bosma}, J., {et~al.} 2020, Nat. Astron., 4,
  913, \dodoi{https://doi.org/10.1038/s41550-020-1182-4}

\bibitem[{{Aiello} {et~al.}(2021){Aiello}, {Albert}, \&
  {Garre}}]{KM3NeT:2021_22}
{Aiello}, S., {Albert}, A., \& {Garre}, S.~A. 2021, Eur. Phys. J. C, 81,
  \dodoi{https://doi.org/10.1140/epjc/s10052-021-09187-5}

\bibitem[{{Alexeyev} \& {Alexeyeva}(2008)}]{Alexeyev:2008_32}
{Alexeyev}, E.~N., \& {Alexeyeva}, L.~N. 2008, Astron. Lett., 34, 745,
  \dodoi{https://doi.org/10.1134/S1063773708110030}

\bibitem[{{Avronin} {et~al.}(2019){Avronin}, {Avronin}, {Aynutdinov},
  {Bannash}, {Belolaptikov}, {Brudanin}, {Budnev}, {Domogatsky}, {Doroshenko},
  {Dvornicky}, \& {Dyachok}}]{Avronin:2019_39}
{Avronin}, A.~D., {Avronin}, A.~V., {Aynutdinov}, V.~M., {et~al.} 2019, in
  {Proceedings of 36th International Cosmic Ray Conference PoS(ICRC2019)}.
\newblock \doarXiv{1908.05427}

\bibitem[{{Bhattacharya} {et~al.}(2015){Bhattacharya}, {Enberg}, {Hall Reno},
  \& {Sarcevic}}]{Bhattacharya&Enberg:2015_07}
{Bhattacharya}, A., {Enberg}, E., {Hall Reno}, M., \& {Sarcevic}, I. 2015,
  \jcap, 2015, \dodoi{https://doi.org/10.1088/1475-7516/2015/06/034}

\bibitem[{{Bionta} {et~al.}(1987){Bionta}, {Blewitt}, {Bratton}, {Casper},
  {Cioccio}, {Claus}, {Cortez}, {Crouch}, {Dye}, {Errede}, {Foster},
  {Gajewski}, {Ganezer}, {Goldhaber}, {Haines}, {Jones}, {Kielczewska},
  {Kropp}, {Learned}, {Losecco}, {Matthews}, {Miller}, {Mudan}, {Park},
  {Price}, {Reines}, {Schultz}, {Seidel}, {Shumard}, {Sinclair}, {Sobel},
  {Stone}, {Sulak}, {Svoboda}, {Thornton}, {van der Velde}, \&
  {Wuest}}]{Bionta:1987_13}
{Bionta}, R.~M., {Blewitt}, G., {Bratton}, C.~B., {et~al.} 1987, \prl, 58,
  1494, \dodoi{https://doi.org/ 10.1103/PhysRevLett.58.1494}

\bibitem[{{Bromberg} {et~al.}(2011){Bromberg}, {Nakar}, {Piran}, \&
  {Sari}}]{Bromberg&Nakar&Piran:2011_62}
{Bromberg}, O., {Nakar}, E., {Piran}, T., \& {Sari}, R. 2011.
\newblock \doarXiv{1112.5949}

\bibitem[{Cano {et~al.}(2017)Cano, Wang, Dai, \& Wu}]{Cano:2017_59}
Cano, Z., Wang, S.-Q., Dai, Z.-G., \& Wu, X.-F. 2017, Advances in Astronomy,
  2017, \dodoi{10.1155/2017/8929054}

\bibitem[{{Enberg} {et~al.}(2009){Enberg}, {Hall Reno}, \&
  {Sarcevic}}]{Enberg:2009_02}
{Enberg}, R., {Hall Reno}, M., \& {Sarcevic}, I. 2009, \prd, 79,
  \dodoi{https://doi.org/10.1103/PhysRevD.79.053006}

\bibitem[{Esmaili \& Murase(2018)}]{Esmaili&Murase:2018_56}
Esmaili, A., \& Murase, K. 2018, Journal of Cosmology and Astroparticle
  Physics, 2018, \dodoi{10.1088/1475-7516/2018/12/008}

\bibitem[{He {et~al.}(2018)He, Kusenko, Nagataki, Fan, \& Wei}]{He:2018_60}
He, H.-N., Kusenko, A., Nagataki, S., Fan, Y.-Z., \& Wei, D.-M. 2018, ApJ, 856,
  \dodoi{10.3847/1538-4357/aab360}

\bibitem[{{Hirata} {et~al.}(1987){Hirata}, {Kajita}, {Koshiba}, {Nakahata},
  {Oyama}, {Sato}, {Suzuki}, {Takita}, {Totsuka}, {Kifune}, {Suda}, {Takashi},
  {Tanimori}, {Miyano}, {Yamada}, {Beier}, {Feldscher}, {Kim}, {Mann},
  {Newcomer}, {Van}, {Zhang}, \& {Cortez}}]{Hirata:1987_12}
{Hirata}, K., {Kajita}, T., {Koshiba}, M., {et~al.} 1987, \prl, 58,
  \dodoi{https://doi.org/10.1103/PhysRevLett.58.1490}

\bibitem[{{Ikeda} {et~al.}(2007){Ikeda}, {Takeda}, \& {Vagins}}]{Ikeda:2007_20}
{Ikeda}, M., {Takeda}, A., \& {Vagins}, M.~R. 2007, The Astrophysical Journal,
  669, \dodoi{https://doi.org/10.1086/521547}

\bibitem[{{Kachelriess} \& {Tomas}(2006)}]{Kachelriess:2006_27}
{Kachelriess}, M., \& {Tomas}, R. 2006, \prd, 74, 063009,
  \dodoi{https://doi.org/10.1103/PhysRevD.74.063009}

\bibitem[{Kheirandish \& Murase(2022)}]{Kheirandish:2022_44}
Kheirandish, A., \& Murase, K. 2022.
\newblock \doarXiv{2204.08518}

\bibitem[{{Koers} \& {Wijers}(2007)}]{Koers:2007_08}
{Koers}, H.~B.~J., \& {Wijers}, R.~A.~M.~J. 2007.
\newblock \doarXiv{0711.4791}

\bibitem[{{Li} {et~al.}(2011){Li}, {Leaman}, {Chornock}, {Filippenko},
  {Poznanski}, {Ganeshalingam}, {Wang}, {Modjaz}, {Jha}, {Foley}, \&
  {Smith}}]{Li:2011_09}
{Li}, W., {Leaman}, J., {Chornock}, R., {et~al.} 2011, Monthly Notices of the
  Royal Astronomical Society, 412, 1441–1472,
  \dodoi{https://doi.org/10.1111/j.1365-2966.2011.18160.x}

\bibitem[{Moriya {et~al.}(2018)Moriya, Förster, Yoon, Gräfener, \&
  Blinnikov}]{Moriya:2018_49}
Moriya, T.~J., Förster, F., Yoon, S.-C., Gräfener, G., \& Blinnikov, S.~I.
  2018, MNRAS, 476, 2840, \dodoi{10.1093/mnras/sty475}

\bibitem[{Moriya {et~al.}(2014)Moriya, Maeda, Taddia, Sollerman, Blinnikov, \&
  Sorokina}]{Moriya:2014_48}
Moriya, T.~J., Maeda, K., Taddia, F., {et~al.} 2014, MNRAS, 439, 2917,
  \dodoi{10.1093/mnras/stu163}

\bibitem[{{Murase}(2018)}]{Murase:2018_01}
{Murase}, K. 2018, \prd, 97, \dodoi{https://doi.org/10.1103/PhysRevD.97.081301}

\bibitem[{Murase {et~al.}(2019)Murase, Franckowiak, Maeda, Margutti, \&
  Beacom}]{Murase&Franckowiak:2019_57}
Murase, K., Franckowiak, A., Maeda, K., Margutti, R., \& Beacom, J.~F. 2019,
  ApJ, 874, \dodoi{10.3847/1538-4357/ab0422}

\bibitem[{Murase \& Ioka(2013{\natexlab{a}})}]{Murase:2013_61}
Murase, K., \& Ioka, K. 2013{\natexlab{a}}, Physical Review Letters, 111,
  \dodoi{10.1103/PhysRevLett.111.121102}

\bibitem[{Murase \& Ioka(2013{\natexlab{b}})}]{Murase:2013_54}
---. 2013{\natexlab{b}}, Physical Review Letters, 111,
  \dodoi{10.1103/PhysRevLett.111.121102}

\bibitem[{Murase {et~al.}(2006)Murase, Ioka, Nagataki, \&
  Nakamura}]{Murase:2006_51}
Murase, K., Ioka, K., Nagataki, S., \& Nakamura, T. 2006, ApJ, 651,
  \dodoi{10.1086/509323}

\bibitem[{Murase {et~al.}(2011)Murase, Thompson, Lacki, \&
  Beacom}]{Murase:2011_47}
Murase, K., Thompson, T.~A., Lacki, B.~C., \& Beacom, J.~F. 2011, Phys.Rev.D,
  84, \dodoi{10.1103/PhysRevD.84.043003}

\bibitem[{{Nakamura} {et~al.}(2016){Nakamura}, {Horiuchi}, {Tanaka}, {Hayama},
  {Takiwaki}, \& {Kotake}}]{Nakamura:2016_28}
{Nakamura}, K., {Horiuchi}, S., {Tanaka}, M., {et~al.} 2016, \mnras, 416,
  3296–3313, \dodoi{https://doi.org/10.1093/mnras/stw1453}

\bibitem[{{Necker}(2021)}]{Necker:2021_31}
{Necker}, J. 2021, in {Proceedings of 37th International Cosmic Ray Conference
  PoS(ICRC2021)}, Vol. 395, 1116, \dodoi{https://doi.org/10.22323/1.395.1116}

\bibitem[{{Nordin} {et~al.}(2019){Nordin}, {Brinnel}, {van Santen}, {Bulla},
  {Feindt}, {Franckowiak}, {Fremling}, {Gal-Yam}, {Kowalski}, {Mahabal},
  {Miranda}, {Rauch}, {Rigault}, {Schulze}, {Reusch}, {Sollerman}, {Stein},
  {Yaron}, {van Velzen}, \& {Ward}}]{Nordin:2019_29}
{Nordin}, J., {Brinnel}, V., {van Santen}, J., {et~al.} 2019, \aap, 631, 14,
  \dodoi{https://doi.org/10.1051/0004-6361/201935634}

\bibitem[{{Oyama}(2022)}]{Oyama:2022_37}
{Oyama}, Y. 2022, \apj, 925, \dodoi{https://doi.org/10.3847/1538-4357/ac4269}

\bibitem[{{Perley} {et~al.}(2020){Perley}, {Fremling}, {Sollerman}, {Miller},
  {Dahiwale}, {Sharma}, {Bellm}, {Biswas}, {Brink}, {Bruch},
  {et~al.}}]{ZTF:2020_41}
{Perley}, D.~A., {Fremling}, C., {Sollerman}, J., {et~al.} 2020, \apj, 904, 35,
  \dodoi{https://doi.org/10.3847/1538-4357/abbd98}

\bibitem[{{Piran} {et~al.}(2019){Piran}, {Nakar}, {Mazzali}, \&
  {Pian}}]{Piran:2019_11}
{Piran}, T., {Nakar}, E., {Mazzali}, P., \& {Pian}, E. 2019, \apjl, 2.
\newblock \doarXiv{1704.08298}

\bibitem[{{Razzaque} {et~al.}(2004){Razzaque}, {Meszaros}, \&
  {Waxman}}]{Razzaque&Meszaros&Waxman:2004_18}
{Razzaque}, S., {Meszaros}, P., \& {Waxman}, E. 2004, \prl, 93,
  \dodoi{10.1103/PhysRevLett.93.181101}

\bibitem[{{Razzaque} {et~al.}(2005){Razzaque}, {Meszaros}, \&
  {Waxman}}]{Razzaque&Meszaros&Waxman:2005_10}
---. 2005, Mod. Phys. Lett. A, 20, 2351,
  \dodoi{https://doi.org/10.1142/S0217732305018414}

\bibitem[{{Rozwadowska} {et~al.}(2021){Rozwadowska}, {Vissani}, \&
  {Cappellaro}}]{Rozwadowska:2021_19}
{Rozwadowska}, K., {Vissani}, F., \& {Cappellaro}, E. 2021, \na, 83,
  \dodoi{https://doi.org/10.1016/j.newast.2020.101498}

\bibitem[{Sarmah {et~al.}(2022)Sarmah, Chakraborty, Tamborra, \&
  Auchettl}]{Sarmah&Tamborra:2022_45}
Sarmah, P., Chakraborty, S., Tamborra, I., \& Auchettl, K. 2022, Journal of
  Cosmology and Astroparticle Physics, 2022, 011,
  \dodoi{10.1088/1475-7516/2022/08/011}

\bibitem[{{Senno} {et~al.}(2016){Senno}, {Murase}, \&
  {Meszaros}}]{Senno&Murase:2016_26}
{Senno}, N., {Murase}, K., \& {Meszaros}, P. 2016, \prd, 93, 083003,
  \dodoi{https://doi.org/10.1103/PhysRevD.93.083003}

\bibitem[{Senno {et~al.}(2018)Senno, Murase, \&
  Meszaros}]{Senno&Murase:2018_55}
Senno, N., Murase, K., \& Meszaros, P. 2018, Journal of Cosmology and
  Astroparticle Physics, 2018, \dodoi{10.1088/1475-7516/2018/01/025}

\bibitem[{Shappee {et~al.}(2014)Shappee, Prieto, Grupe,
  {et~al.}}]{ASASSN:2014_62}
Shappee, B., Prieto, J., Grupe, D., {et~al.} 2014, ApJ, 788,
  \dodoi{10.1088/0004-637X/788/1/48}

\bibitem[{Shin'ichiro \& Beacom(2005)}]{Shinichiro&Beacom:2005_50}
Shin'ichiro, A., \& Beacom, J.~F. 2005, Physical Review Letters, 95,
  \dodoi{10.1103/PhysRevLett.95.061103}

\bibitem[{{Zirakashvili} \& {Ptuskin}(2016)}]{Zirakashvili&Ptuskin:2016_25}
{Zirakashvili}, V.~N., \& {Ptuskin}, V.~S. 2016, Astroparticle Physics, 78, 28,
  \dodoi{https://doi.org/10.1016/j.astropartphys.2016.02.004}

\end{thebibliography}
\bibliographystyle{aasjournal}

\end{document}